\documentclass[aps,prb,preprint,groupedaddress,showpacs]{revtex4}

\usepackage{graphicx}
\usepackage{dcolumn}
\usepackage{bm}
\usepackage{subfigure}

\begin{document}


\title{Structure Factor of the $3D$ Random Field $XY$ Model}


\author{Ronald Fisch}
\email[]{ron@princeton.edu}
\affiliation{382 Willowbrook Dr.\\
North Brunswick, NJ 08902}


\date{\today}

\begin{abstract}
We have performed Monte Carlo studies of the 3D random field $XY$
model on $L \times L \times L$ simple cubic lattices, with random
field strengths of $h_r$ = 1 and 2.  We present results for the
angle-averaged magnetic structure factor, $S ( k )$ at $L = 64$.
Our results appear to indicate a phase transition into a
ferromagnetic state.  This is made possible by the existence of a
Griffiths singularity.  It appears that at the phase transition
$M^2$ jumps to zero discontinuously, with a latent heat which is
probably subextensive.

\end{abstract}


\maketitle

\section{Introduction}

Some time ago, Monte Carlo calculations\cite{GH96,Fis97} indicated
that there appeared to be a region of quasi-long-range order (QLRO)
in the phase diagram of the three-dimensional (3D) random-field $XY$
model (RFXYM), at weak random field and low temperature.  Some
recent functional renormalization group calculations\cite{DW06,TT06}
have questioned this.  Therefore, since there have been substantial
improvements in computing power over the last ten years, the author
felt it worthwhile to conduct a new Monte Carlo study of this model.

For fixed-length classical spins the Hamiltonian of the RFXYM is
\begin{equation}
  H ~=~ - J \sum_{\langle ij \rangle} \cos ( \phi_{i} - \phi_{j} )
  ~-~ h_r \sum_{i} \cos ( \phi_{i} - \theta_{i} )  \, .
\end{equation}
Each $\phi_{i}$ is a dynamical variable which takes on values
between 0 and $2 \pi$. The $\langle ij \rangle$ indicates here a sum
over nearest neighbors on a simple cubic lattice of size $L \times L
\times L$. We choose each $\theta_{i}$ to be an independent
identically distributed quenched random variable, with the
probability distribution
\begin{equation}
  P ( \theta_i ) ~=~ 1 / 2 \pi   \,
\end{equation}
for $\theta_i$ between 0 and $2 \pi$.  This Hamiltonian is closely
related to models of vortex lattices and charge density waves.

Larkin\cite{Lar70} studied a model for a vortex lattice in a
superconductor.  His model replaces the spin-exchange term of the
Hamiltonian with a harmonic potential, so that each $\phi_{i}$ is no
longer restricted to lie in a compact interval.  He argued that for
any non-zero value of $h_r$ this model has no ferromagnetic phase on
a lattice whose dimension $d$ is less than or equal to four.  A more
intuitive derivation of this result was given by Imry and
Ma,\cite{IM75} who assumed that the increase in the energy of an
$L^d$ lattice when the order parameter is twisted at a boundary
scales as $L^{d - 2}$.

As argued by Imry and Ma,\cite{IM75} and later justified more
carefully,\cite{AIM76,PS82} within a perturbative $\epsilon$
expansion one finds the phenomenon of ``dimensional reduction". The
critical exponents of any $d$-dimensional $O(N)$ random-field model
(where $N$ is the number of components of the spin on each site)
appear to be identical to those of an ordinary $O(N)$ model of
dimension $d - 2$. For the Ising ($N = 1$) case, this was shown
rigorously to be incorrect.\cite{Imb84,BK87}

Because translation invariance is broken for any non-zero $h_r$, it
seems quite implausible that the twist energy for Eqn.~(1) scales
as $L^{d - 2}$, even though this is correct to all orders in
perturbation theory.  An alternative derivation by Aizenman and
Wehr,\cite{AW89,AW90} which claims to be mathematically rigorous,
also makes an assumption equivalent to translation invariance.
Although the average over the probability distribution of random
fields restores translation invariance, one must take the infinite
volume limit first.  It is not correct to interchange the infinite
volume limit with the average over random fields.

The basic point is that the Imry-Ma argument for continuous spins
($O ( N )$ with $N \ge 2$) is not self-consistent.  One begins by
assuming that the random field is weak so that the twist energy
scales as $L^{d - 2}$, as in the absence of the random field.  Then
one shows that if $d \le 4$, this cannot be true for large enough
$L$.  The only conclusion which should be drawn from this is that a
deeper analysis of what is going on is needed in that case.

\section{The Griffiths phase}

A mechanism which causes the breakdown of perturbation theory in the
presence of quenched randomness at the critical temperature, $T_c$,
of the pure ferromagnet was discovered by Griffiths\cite{Grif69} in
the special case of a bond-diluted Ising model.  It was later shown
explicitly\cite{Har74} that the magnetization per unit volume,
$\vec{\bf M} ( \vec{\bf H} )$ has an essential singularity at
$\vec{\bf H} = 0$ (where $\vec{\bf H}$ is a uniform field) at and
below $T_c$. The result was then extended to the case of binary
ferromagnetic alloys.\cite{Fis81}  It was later argued that this
result could be extended beyond the Ising model to $O ( N )$
models,\cite{Dot99} and that the Griffiths singularity will exist
whenever the specific heat is divergent at $T_c$ for the pure model,
so that the bond randomness is a relevant perturbation.\cite{Har74b}

This argument appears to imply that it is never correct to use a
perturbation theory of the standard type ({\it e.g.} an $\epsilon$
expansion) when the randomness is a relevant perturbation.  On the
other hand, an $\epsilon$ expansion is only an asymptotic series
even in the nonrandom case.  Thus one may hope that it might
continue to be useful even in the presence of the Griffiths
singularity.

The original concept of Griffiths\cite{Grif69} was a singularity
caused by the contributions of large, isolated clusters of spins.
Hertz, Fleishman and Anderson\cite{HFA79} developed a more intuitive
picture of the Griffiths singularity in random bond models, which
was developed further by Bray and Moore\cite{BM82,Bra87}.  These
authors showed that the same type of singular behavior would be a
consequence of a Lifshits tail of localized eigenvectors of the
inverse magnetic susceptibility matrix, $\chi^{-1}$, near the band
edge at eigenvalue zero.

If one generalizes the idea of the Griffiths singularity from
these random bond models to random fields models, one finds
similarities, but also important differences.  The magnetic
structure factor, $S (\vec{\bf k}) = \langle | \vec{\bf M}(\vec{\bf
k}) |^2 \rangle $, for $XY$ spins is
\begin{equation}
  S (\vec{\bf k}) ~=~  L^{-3} \sum_{ i,j } \cos ( \vec{\bf k} \cdot
  \vec{\bf r}_{ij}) \langle \cos ( \phi_{i} - \phi_{j}) \rangle  \,   ,
\end{equation}
where $\vec{\bf r}_{ij}$ is the vector on the lattice which starts
at site $i$ and ends at site $j$, and here the angle brackets denote
a thermal average.  For a random field model, unlike a random bond
model, the longitudinal part of the magnetic susceptibility, $\chi$,
which is given by
\begin{equation}
  T \chi (\vec{\bf k}) ~=~ 1 - M^2 ~+~ L^{-3} \sum_{ i \ne j } \cos (
  \vec{\bf k}  \cdot \vec{\bf r}_{ij}) (\langle \cos ( \phi_{i} - \phi_{j}
  ) \rangle ~-~ Q_{ij} )  \,   ,
\end{equation}
is not the same as $S$ even above $T_c$.  For $XY$ spins,
\begin{equation}
  Q_{ij} ~=~ \langle \cos ( \phi_{i} ) \rangle \langle \cos (
  \phi_{j} ) \rangle ~+~ \langle \sin ( \phi_{i} ) \rangle \langle
  \sin ( \phi_{j} ) \rangle  \,  ,
\end{equation}
and
\begin{equation}
  M^2 ~=~ L^{-3} \sum_{i} Q_{ii}
      ~=~ L^{-3} \sum_{i} \langle \cos ( \phi_{i} ) \rangle^2 ~+~
       \langle \sin ( \phi_{i} ) \rangle^2  \,  .
\end{equation}
When there is a ferromagnetic phase transition, $S ( \vec{\bf k} = 0
)$ has a stronger divergence than $\chi ( \vec{\bf k} = 0 )$. Thus
it is $S$, and not $\chi$, which is the best place to look for an
essential singularity in the random field model.  There may also be
an effect\cite{BM82} in $\chi$, but we should look in the place
where the effect is expected to be largest.

The author is not aware of any explicit studies on the question of
the occurrence of Lifshits tails in a matrix which has the form of
$S (\vec{\bf k})$.  The natural generalization of the Harris
criterion is that the Griffiths singularity should be found in $S (
\vec{\bf k} = 0)$ at the $T_c$ of the pure $O ( N )$ ferromagnet for
any finite $N$, and any $d > 2$.  This is because the response of
the pure system to a weak uniform $\vec{H}$ field is always
divergent at $T_c$, {\it i.e.} the susceptibility exponent $\gamma$
is always positive.\cite{SA81}

A detailed numerical study of the 3D Ising model in a random field
(RFIM) at $T = 0$ has been performed by Middleton and
Fisher.\cite{MF02}  These authors find that the 3D RFIM has a
ferromagnetic critical point, and that the order parameter exponent
$\beta$ has a small, but positive value.  For weak random fields,
the 3D RFIM still has two low-temperature (ferromagnetic) Gibbs states,
even though the random fields have destroyed the Kramers degeneracy.
The current author believes the existence of multiple ferromagnetic
Gibbs states is required for the existence of a critical point with
a positive $\beta$.

Middleton and Fisher find no explicit evidence for the existence of a
Griffiths phase in the RFIM, although they agree that such a phase should
exist.  Middleton\cite{Mid02} has also studied the 4D RFIM at $T = 0$.

The scalar quantity $\langle M^2 \rangle$ is a well-defined function
of the lattice size $L$ for finite lattices, which, with high
probability, approaches its large $L$ limit smoothly as $L$ increases.
The vector $\vec{\bf M}$, on the other hand, is not really a
well-behaved function of $L$ for an $XY$ model in a random field.
Knowing the local direction in which $\vec{\bf M}$ is pointing,
averaged over some small part of the lattice, may not give us a strong
constraint on what $\langle \vec{\bf M} \rangle$ for the entire
lattice will be.

\section{Numerical results for $S ( k )$}

In this work, we will present results for the average over angles of
$S (\vec{\bf k})$, which we write as $S ( k )$.  The data were
obtained from $L \times L \times L$ simple cubic lattices with $L =
64$ using periodic boundary conditions.  Some preliminary studies
for smaller values of $L$ were also done.  The calculations were
done using a 12-state clock model, {\it i.e.} a ${\bf Z}_{12}$
approximation\cite{Fis97} to the $XY$ model of Eqn.~(1).  The
strengths of the random field for which data were obtained are $h_r$
= 1 and 2.  These sets of parameters allow the use of a lookup table
for the Boltzmann factors, because all the energies in the problem
are expressible as sums of integers and integer multiples of
$\sqrt{3}$.\cite{Note1}

This discretization of the phase space of the model has significant
effects at very low $T$, but the effects at the temperatures we
study here are expected to be negligible compared to our statistical
errors.  The probability distributions for the local magnetization
of equilibrium states which are calculated for the ${\bf Z}_{12}$
model are found to have very small contributions from the third and
higher harmonics of $\cos ( \phi )$ and $\sin ( \phi )$.  This
is strong evidence that the 12-state clock model is an accurate
approximation to the $XY$ model within our range of parameters.

The program uses two linear congruential pseudorandom number
generators, one for choosing the values of the $\theta_i$, and a
different one for the Monte Carlo spin flips, which are performed by
a single-spin-flip heat-bath algorithm.  The code was checked by
setting $h_r = 0$, and seeing that the expected behavior of the pure
ferromagnetic system was produced correctly.

\begin{figure}
\includegraphics[width=3.4in]{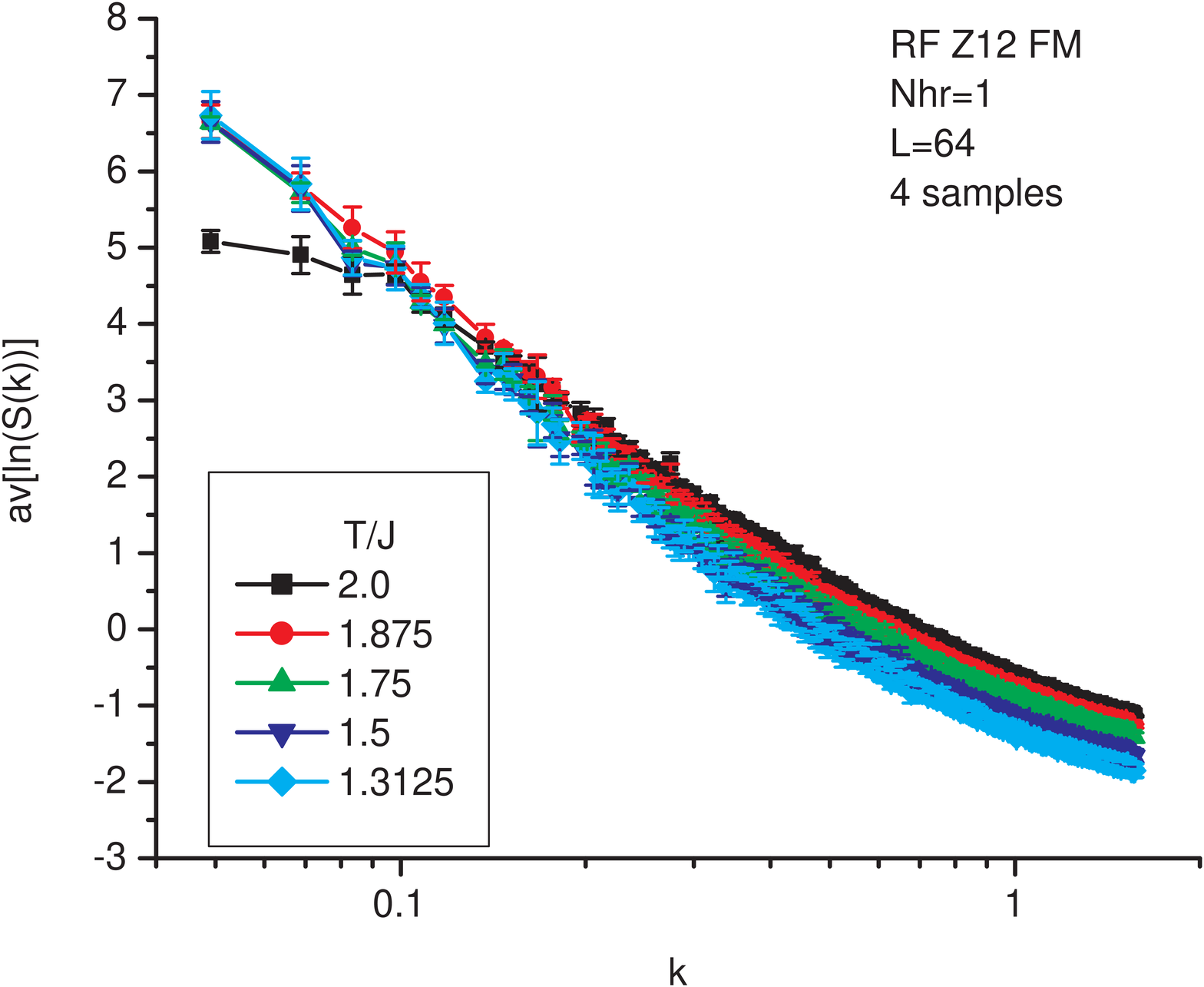}
\caption{\label{Fig.1}(color online) Angle-averaged structure factor
for $64 \times 64 \times 64$ lattices with $h_r = 1$ at various
temperatures. The error bars indicate one standard deviation
statistical errors, and the $x$-axis is scaled logarithmically.}
\end{figure}

Four different $L = 64$ realizations of the random fields $\theta_i$
were used.  The same four samples of random fields were used for all
values of $T$, and both values of $h_r$.  Each lattice was started
off in a random spin state at $T / J = 2.3125$, significantly above
the $T_c$ for the pure model, and cooled slowly.  Thermal averages
for $S (\vec{\bf k})$ were obtained at a set of temperatures.  At
each $T$ a sequence of 12 spin states obtained at intervals of
20,480 Monte Carlo steps per spin (MCS) was Fourier transformed and
averaged. The data were then binned according to the value of $k^2$,
to give the angle-averaged $S ( k )$.  Finally, a logarithically
weighted average over the four samples was performed.

The data for $h_r = 1$ at $T / J$ between 1.3125 and 2.0 are shown
in Fig.~1.  At $T / J = 2.0$, $S$ is clearly flattening out for $k <
0.1$.  As $T$ is lowered, weight is shifted from large values of $k$
toward $k = 0$.  For all the smaller values of $T / J$ shown in the
figure, the correlation length, $\xi$, is clearly larger than the
size of the samples, $L = 64$. All of these large $\xi$ samples show
an $A k^{-3}$ behavior at small $k$, and the coefficient $A$ is,
within our statistical errors, independent of $T$.  This apparently
universal behavior is consistent with the prediction of Giamarchi
and Le Doussal\cite{GD95} for the 3D elastic glass model. For the
elastic glass model, this $A k^{-3}$ behavior has be verified
numerically by McNamara, Middleton and Zeng.\cite{MMZ99}  For the
RFXYM, however, we know that this behavior cannot hold down to $k =
0$, because the fixed length of the $XY$ spins yields a sum rule for
$S$:
\begin{equation}
  \sum_{\vec{\bf k}} S (\vec{\bf k}) ~=~  L^{3}   \,   .
\end{equation}

\begin{figure}
\includegraphics[width=3.4in]{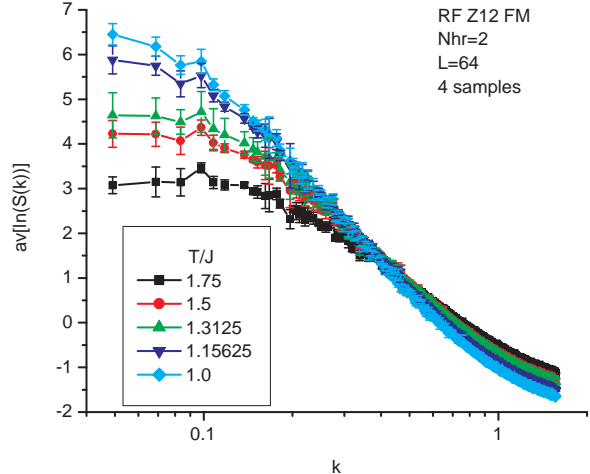}
\caption{\label{Fig.2}(color online) Angle-averaged structure factor
for $64 \times 64 \times 64$ lattices with $h_r = 2$ at various
temperatures.  The error bars indicate one standard deviation
statistical errors, and the $x$-axis is scaled logarithmically.}
\end{figure}

There is also some interesting information to be learned from the
dynamical behavior.  Below $T / J = 1.875$, one of the four samples
shows strong hysteresis, and a second one shows mild hysteresis.
This may indicate the existence of a first-order phase transition
for $h_r = 1$ at $T / J \approx 1.80$.  The hysteresis goes away as
we continue to still lower $T$.  The value of $\xi$ under these
conditions is clearly bigger than $L = 64$, however.  Therefore we
cannot say exactly what is going on here from these data.

\begin{figure}
\includegraphics[width=3.4in]{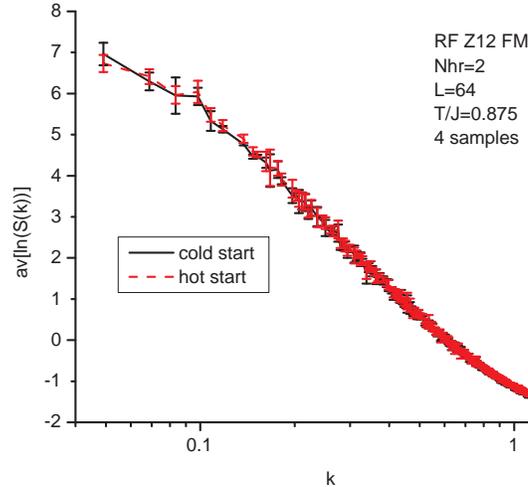}
\caption{\label{Fig.3}(color online) Angle-averaged structure factor
for $64 \times 64 \times 64$ lattices with $h_r = 2$ at $T/J$ =
0.875, using both hot start and ordered start initial conditions.
The error bars indicate one standard deviation statistical errors,
and the $x$-axis is scaled logarithmically.}
\end{figure}

Due to the sum rule, Eqn.~(7), we know that the $A k^{-3}$ behavior
which is seen in Fig.~1 {\em cannot} be the correct behavior at
small $k$ for very large lattices.  Increasing the size of our
lattice by a significant amount is rather impractical.  However, it
is impossible to distinguish between a ferromagnetic state and the
QLRO state claimed in the earlier Monte Carlo
studies\cite{GH96,Fis97} without studying the true small $k$ limit.

Another way to learn more about the behavior is to decrease $\xi$ by
increasing $h_r$.  In Fig.~2, we display $S ( k )$ data from four
$h_r = 2$ samples, obtained using the same procedures as before.  In
this case the peak at $\vec{\bf k} = 0$ continues to grow as we
lower $T$, but there does not appear to be any simple scaling
behavior.  The peaks are very flat near $\vec{\bf k} = 0$.  An
attempt to fit the data by a Lorentzian shape, or even a Lorentzian
raised to some power, is not satisfactory.

At $T / J = 0.875$, a system which is started in an initially
ordered state shows no sign of relaxing into the low $M^2$ state
which is found by slow cooling, even on time scales of $10^6$ MCS
per spin. In Fig.~3 we show $\langle \ln ( S ( k )) \rangle$,  with
$h_r = 2$, at this temperature, averaged over the same 4 samples for
both types of initial conditions.  From this figure we see that,
although the two initial conditions give significantly different
values of $\langle M^2 \rangle$ the rest of $\langle \ln ( S ( k
)) \rangle$ is almost the same.  The peak in $\langle \ln ( S ( k ))
\rangle$  has continued to sharpen slowly with the reduction in $T$.
The values of $\langle \ln ( S ( 0 )) \rangle$, not shown in Fig.~3,
are $11.54 \pm 0.11$ for the initially ordered state, and $8.26 \pm
0.28$ for the slow-cooled state.

The ordered initial conditions give slightly lower values of the
energy per spin, as we will discuss in more detail in the next
section.  A careful study to find the highest value of $T$ for which
the metastability exists was not performed.  For $h_r = 2$, unlike
$h_r = 1$, the relaxation time gets increasingly longer as we
continue to lower $T$ beyond the point where metastability first
appears.

\section{Discussion}

If, as argued by Imry and Ma, the random-field energy dominates the
spin-exchange energy for the 3D RFXYM, how could it be possible to
have a ferromagnetic phase? To understand this, we need to think
carefully about finite-size scaling.

Consider a very large system at $T = 0$, with $h_r = 1$ (for
example).  Since there is no degeneracy due to symmetry, and the
energy is not quantized, we expect that the ground state for any
particular sample of random fields should be unique, with
probability one.  If $d > 4$, then we expect that $S ( \vec{\bf k})$
has a $\delta$-function peak at $\vec{\bf k} = 0$, of strength $M^2
( T = 0 ) > 0$.  If we consider some low, but finite $T$, we expect
that, in thermodynamic equilibrium, many states will contribute to
$S ( \vec{\bf k})$.  These states should have similar values of
$M^2 > 0$, but the direction of $\vec{\bf M}$ ought to be randomly
distributed on the unit circle, so that $\langle \vec{\bf M} \rangle
\approx 0$.

In $d = 3$, the Monte Carlo results presented here do not indicate
that there are many states with similar values of $M^2 > 0$, but
very different values of the direction of $\vec{\bf M}$,
contributing substantially to the equilibrium Gibbs state.  This may
happen as a rare event in a few samples, but it is unlikely to be
observed. There is, however, an alternative road to ferromagnetism.

Let us consider some very large, but finite, 3D sample.  As Bray
has argued,\cite{Bra87} $\chi^{-1}$ has a Lifshits tail of localized
states near eigenvalue 0.  These localized states occur in regions
of the sample in which the average local random field has a large
magnitude, but a random direction.  At any temperature, each such
localized state corresponds to a compact cluster of spins which is
pinned to point in a direction which is approximately parallel to
the local field in that neighborhood.

When we lower $T$, the eigenvalue spectrum of $\chi^{-1}$ will be
modified, as discussed by Hertz, Fleishman and Anderson.\cite{HFA79}
For $d > 4$ mobility edge will move down to zero, and the sample
polarizes into the renormalized band-edge state.  In 3D this type of
continuous process cannot occur.  However, it may become possible,
at a low enough $T$, for the state of minimum eigenvalue to polarize
most of the sample, without completely destroying the other stable
clusters.

A condition which makes it favorable for this process to occur is
that the density of localized states should be low, and that the
size of a typical localized state should be small compared to the
average nearest neighbor distance between the localized states.
Whether this is possible may thus depend on the details of the
distribution of random fields.  An earlier calculation\cite{Fis97}
using a different type of random field distribution provided
evidence for a continuous phase transition.

A small difference in energy per spin,
\begin{equation}
\langle \Delta E \rangle ~=~ (-1.24 \pm 0.25)\times 10^{-3}~J  \,
\end{equation}
per spin, was found in our $L = 64$ results for $h_r = 2$ at $T / J =
0.875$ between the slowly cooled states and the states relaxed from
ordered initial conditions.  In order to reach this low energy state,
the direction of the initial $\vec{M}$ must be close to the direction
of $\langle \vec{M} \rangle$ for the slow-cooled state.  Using an
initial $\vec{M}$ which is not close to this direction results in a
state which has a higher energy than the slow-cooled state.

The author believes that this energy is subextensive.  This means
that, as $L \to \infty$, $\langle \Delta E \rangle \to 0$, but $L^3
\langle \Delta E \rangle \to \infty$.  In these terms, the energy
difference quoted above is $-325 \pm 66 ~J$.  Thus the Boltzmann
factor, $\exp ( -325 / 0.875 )$, for the weighting of the two
separate free energy minima is extremely small.  Since, of course,
the free energies of the two Gibbs states must be equal at the phase
transition, this Boltzmann factor must be compensated by the entropy
difference between them.

It has been known for a long time that something similar
occurs\cite{Tho69,AY71} in the 1D Ising model with inverse-square
law long-range interactions.  For this model it is known that the
correlation length diverges at $T_c$,\cite{BCRS81} so that
$T_c$ is a critical point with $\beta = 0$.  A phase transition of
this type is also believed to occur in the {\it k}-core percolation
model,\cite{SLC06} which has been suggested as model for the
ordinary glass transition.

A rather similar subextensive singularity was recently seen in a type
of 2D Ising spin glass.\cite{Fis07}  If this effect also occurs in
the Ising spin glass in higher dimensions, where $T_c > 0$, it
provides an explanation for the puzzling behavior of the high
temperature series for the Ising spin glass,\cite{FH77} since we know
that for random bond Ising models the Griffiths singularity appears
in 4D.  This is consistent with the ideas of Bray and
Moore.\cite{BM82,Bra87}

A phase transition at about $T / J \approx 1.80$ for $h_r = 1$ and
$T / J \approx 0.90$ for $h_r = 2$ is in good agreement with the work
of Gingras and Huse,\cite{GH96} who worked at $T / J = 1.5$, and
estimated the transition to be at $h_r \approx 1.35$.

\section{Summary}

In this work we have performed Monte Carlo studies of the 3D RFXYM
on $L = 64$ simple cubic lattices, with random field strengths of
$h_r$ = 1 and 2.  We present results for the structure factor,
$S ( k )$, at a sequence of temperatures.  We argue that our results
appear to indicate a phase transition into a ferromagnetic state.
This is made possible by the existence of a Griffiths singularity,
which invalidates the Imry-Ma analysis.  At the phase transition
$M^2$ seems to jump to zero discontinuously, but with a latent heat
per spin which probably goes to zero as $L \to \infty$.

\begin{acknowledgments}
The author thanks P. W. Anderson and D. A. Huse for helpful
discussions, and the Physics Department of Princeton University for
providing use of computers on which the data were obtained.

\end{acknowledgments}



\end{document}